\documentclass[amsmath, amssymb, aps, reprint]{revtex4-2}

\usepackage{titlesec}
\titlespacing*{\subsection}
{0pt}{\baselineskip}{\baselineskip}

\usepackage{graphicx}
\usepackage{amsmath}
\usepackage{bm}
\usepackage{textcomp}
\usepackage[english]{babel}
\usepackage[utf8]{inputenc}
\usepackage{times}
\usepackage{cancel}
\usepackage[T1]{fontenc}
\usepackage{color}
\usepackage{here}
\usepackage{amsfonts}
\usepackage[percent]{overpic}
\usepackage{soul}
\usepackage{gensymb}

\DeclareMathAlphabet{\mathpzc}{OT1}{pzc}{m}{it}

\renewcommand{\hat}[1]{\bm{\widehat{#1}}}

\newcommand{\update}[1]{\textcolor{black}{#1}}

\begin{document}

\title{Light-responsive active particles in a thermotropic liquid crystal}

\author{Antonio Tavera-V\'{a}zquez\textsuperscript{1}}
\thanks{These authors contributed equally to this work}
\author{Andr\'{e}s~C\'{o}rdoba\textsuperscript{1}}
\thanks{These authors contributed equally to this work}
\author{Sam Rubin\textsuperscript{1}}
\author{\textcolor{black}{Vincenzo Vitelli\textsuperscript{2,3}}} 
\author{Juan J. de Pablo\textsuperscript{1,4}}
\thanks{To whom correspondence should be addressed.\\
            Email: depablo@uchicago.edu}

\affiliation{
    \textsuperscript{1}Pritzker School of Molecular Engineering, University of Chicago, 5640 South Ellis Avenue Chicago, IL 60637, USA} 
\affiliation{
    \textcolor{black}{\textsuperscript{2}James Franck Institute and Department of Physics, University of Chicago, Chicago, IL 60637, USA}}
\affiliation{
    \textcolor{black}{\textsuperscript{3}Kadanoff Center for Theoretical Physics, University of Chicago, Chicago, IL 60637, USA}}
\affiliation{
    \textsuperscript{4}Materials Science Division, Argonne National Laboratory, Lemont, IL 60439, USA}


\begin{abstract}
The development of synthetic microswimmers has advanced our understanding of the fundamental self-propelled mechanisms of living systems. However, there are scarce studies at the microscale within highly structured anisotropic media, such as bacteria or cellular receptors that swim in concentrated solutions of filamentous proteins or lipids with viscoelastic properties. Synthetic liquid crystals (LCs) have the potential to 
serve as biomimetic surrogates to study the structure and dynamics 
of living systems. Nevertheless, studies on thermotropic LCs have mainly focused on electro and magneto-phoretic effects, with a few others on diffusiophoresis or light-driven distortions of the LC nematic director.
To the best of our knowledge, here we report self-thermophoretic experiments on thermotropic LCs for the first time. Our system consists of 2D confined Janus particles in 5CB with homeotropic anchoring on the particle and LC cell surfaces. The Janus particles include a conductive titanium coating that, upon exposure to an LED source, is heated and induces a local steady nematic-isotropic phase transition, leading to the self-propulsion of the particles orthogonally to the LC director. The trajectories of the Janus particles were tracked at different intensities of the applied light. A model is developed to describe the mean-squared displacement of a Janus particle suspended in a nematic LC. The model assumes that the Janus particle feels the LC as a continuum with the anisotropic viscosity of the bulk nematic phase. Moreover, the viscoelasticity of the LC is also considered. The model describes the experimental data well, and the fitting parameter related to the magnitude of the swimming force increases with the intensity of the applied light. \textcolor{black}{This analysis suggests the possibility of utilizing active Janus particles for microrheological measurements within nematic LCs. 
Moreover, our approach represents an important step for developing a platform for highly structured anisotropic active materials.}
\end{abstract}


\maketitle

\section{INTRODUCTION}\label{introduction}

Swimming microorganisms, which include bacteria, algae, and spermatozoa, 
are ubiquitous in most biological processes. 
These biological swimmers continuously 
convert local energy into propulsive forces, which allows them to 
move through their surrounding fluid medium much faster than 
by simple diffusion.  
Similarly, synthetic colloidal microswimmers are capable of mimicking complex 
biolocomotion employing simple self-propulsion mechanisms. 
Artificial self-propelled micro- and nano-engines, or swimmers, have been
increasingly attracting the interest of experimental and theoretical researchers.
Advances in microscopy techniques 
have allowed synthetic microswimmers to be studied experimentally
\cite{EbbensLangmuir2011, gomez2017tuning, Michelin2015, Wilson2013}. 
Janus particle swimmers made by
coating fluorescent polymer beads with hemispheres of platinum 
have been fully characterized using video microscopy to
reveal that they undergo propulsion in hydrogen peroxide fuel
away from the catalytic platinum patch \cite{EbbensLangmuir2011}.
Recently, silica microparticles
with a copper catalytic patch were also shown to undergo propulsion in 
hydrogen peroxide fuel, but
towards their catalytic patch \cite{sharan2022upstream}.
The two examples cited above are cases of artificial microswimmers 
where swimming is driven by a local self-generated concentration 
gradient, usually referred to as self-diffusiophoresis. Other works have 
reported artificial swimmers that utilize the 
mechanism known as self-thermophoresis. For example
silica colloidal particles half-coated with gold have been observed
under laser irradiation \cite{jiang2010active}.  
Absorption of a laser at the metal-coated side of the particle creates
local temperature gradient, which in turn drives the particle by thermophoresis.
Moreover, other recent studies of self-propulsion of half-coated spherical colloids in 
critical binary mixtures have shown that the coupling of local body forces, 
induced by laser illumination, and the
wetting properties of the colloid can be used to finely 
tune both the colloid's swimming speed and its directionality \cite{gomez2017tuning}.

Microswimmers have also been studied theoretically and through computer simulations 
\cite{Datt2018, Yasuda2017, Nasouri2017, Milster2017, 
childressJFM2012,sabass2012dynamics, FalascoPRE2016, 
golestanian2005propulsion,
popescu2010phoretic,thakur2011dynamics,cordoba2020simple}. 
For instance, self-propelled Janus particles
that catalyze a chemical reaction inside the fluid have been studied extensively using simulations. 
It has been shown that the catalytic reaction produces an asymmetric, 
non-equilibrium distribution of reaction products
around the colloid, which generates osmotic or other
phoretic forces \cite{golestanian2005propulsion, popescu2010phoretic, 
sabass2012dynamics,thakur2011dynamics, LugliJPCC2012}. 
From a theoretical perspective, a fundamental challenge when 
describing active matter is to properly describe the coupling 
between fuel consumption and the generation of propulsive forces.
Recently, a thermodynamically compliant coupling between fuel consumption 
and motion was proposed for a model of microswimmers driven by self-diffusiophoresis
\cite{cordoba2020simple}. Predictions from that theory indicate that the
time scales at which swimming motion is observed
increases exponentially with the swimmers' length and 
decreases exponentially with the ratio between the diffusivity 
of product and reactant. Moreover,
reactions in which the product has a larger diffusion
coefficient than the reactant produce faster swimming. Although, to a lesser extent, swimming driven by self-thermophoresis has also been studied theoretically. For example, Langevin equations for the self-thermophoretic 
dynamics of Janus particles partially coated with an absorbing layer that is heated 
by a radiation field have been proposed \cite{gaspard2019stochastic}.
In that work, the surface pressure tensor
and heat flux vector were expressed in terms of the slip velocity and 
surface temperature gradient, constructed to be consistent 
with microscopic reversibility.

Most of the previous experimental and theoretical work 
with artificial microswimmers has focused on 
swimming in purely viscous and isotropic fluids. 
However, in nature, it is common for microswimmers, such as bacteria 
or eukaryotic cells to swim 
in fluids where highly concentrated solutions of filamentous proteins such 
as actin or tubulin can form complex phases, with anisotropic and viscoelastic 
properties \cite{spagnolie2023swimming, li2021microswimming}.
In this regard, the filaments that make up the cytoskeleton can experience spatial order and
alignment both at the level of the mesh size ($\simeq$ 10 nm) and of the
whole cell ($\simeq$ 10 $\mu$m), leading to short- and long-range
directionality \citep{luby1999cytoarchitecture,
alvarado2014alignment}. Recent studies have suggested that the anisotropy
of the cytoskeleton plays an important
role in controlling the directionality of important cell functions, such
as mechanotransduction \citep{scherp2007anisotropic, del2008anisotropic}.
Furthermore, even simplified {\it in vitro} reconstituted
models of the cytoskeleton, such as F-actin solutions, exhibit nematic
phases in the bulk at concentrations above 2.5 mg/ml \citep{furukawa1993formation, kas1996f}.
Janus particle swimmers embedded in synthetic nematic liquid crystal (LC)
phases have the potential to be a good biomimetic system for studying the transport 
mechanisms that occur in complex anisotropic biomaterials 
\cite{cordoba2016anisotropy,gomez2013flow,gomez2016two}.

An important body of work has already been conducted
to study the behavior of nano- and microparticles
immersed in nematic LCs. These have led to a growing understanding of 
LC-mediated particle assembly \citep{guzman2003,musevic2006,Tomar2012,amin2015}. 
Less is known, however, about non-equilibrium, dynamic effects \citep{gettelfinger2010flow, koenig2009single, moreno2011effects, stieger2014effects}.
A nematic LC is characterized by a high degree of orientation of the molecules
(mesogens) along a specific direction. Nevertheless, introducing a colloid into an LC can greatly affect the
physical properties of its host and distort the orientational order of the
nematic LC. The distortion is caused by the alignment of the mesogens on the curved
surface of the colloid, and perturbations of the director field lead to long-range
anisotropic forces. It has been shown that flow and particle size have a significant
impact on defect structures around a colloidal particle \citep{stark2001stokes,
yoneya2005effect, araki2006surface, khullar2007dynamic, gettelfinger2010flow, 
stieger2014effects}. The diffusivity of nanoparticles embedded in a LC host has also
been examined using molecular simulations \citep{moreno2011effects}.
Accordingly, the surface anchoring of LC molecules on
the particle does not affect its diffusivity in a simple, monotonic
fashion, but actually depends on the interplay of mesogen-colloid interactions and
mesogen-mesogen ordering in the bulk. For nematic LCs, the diffusivity of the particle is
highly sensitive to the orientation of surface-bound mesogens, attaining an anisotropic anomalous diffusion at short times, and higher mean-squared displacements (MSD) in the direction parallel to the surface-mesogen orientation \cite{LavrentovichScience2013}.
Note that the work discussed above has focused on the passive (or Brownian) motion
of nano- and microparticles in nematic LCs but the active (or ballistic) motion
of microparticles in nematic phases still remains under active exploration.

Previous reports of transport of colloids embedded in thermotropic LCs also exist. The transport of colloids can occur by controlled distortions on the LC director manipulated with external fields or actively altering the colloid-LC interactions at the level of individual particles, giving rise to a local symmetry breaking. Examples of the first case are the electric manipulation of thermotropic LCs when inducing a potential difference transversally to the LC cell confinement. The induced electric field can reorient the LC polar molecules in a direction either parallel \citep{LavrentovichPRL2007} or perpendicular \citep{SMSasaki, SaguesSoftMatter2013} to the field. The specific direction in which the molecules align depends on the sign of the LC dielectric anisotropy. This allows the manipulation of 
colloids that are embedded in the LC but that  are not susceptible to the external field. On the other hand, an example of local active transport of colloids in thermotropic LCs include, ferromagnetic/SU-8 plates that are altered by rotating magnetic fields \citep{StebeScience2022}.
This plates develop characteristic dynamics due to the local emergence of 
disclination lines. Another example is the functionalization of
the surface of silica colloids with azobenzene molecules altering their local conformation under polarized illumination, inducing tunable LC-surface and elastic distortions \citep{SmalyukhNattComm2012}. There is also one recent report of a silica-palladium Janus particle swimming by self-diffusiophoresis in an aqueous-LC interface that shows that the anisotropic viscoelastic environment of LCs leads to new dynamical behaviors of active colloids \cite{mangal2017active}. \textcolor{black}{
Additionally, other approaches to manipulate colloids or 
micrometre-scale impurities embedded in LCs 
utilize temperature gradients 
\cite{samitsu2010molecular,kim2012molecular,vskarabot2013transport,MusevicSoftMatter2017}. }
For example,  the thermoviscous expansion of a thermotropic LC by focalized laser illumination has been studied. The local increment of the temperature melts the LC near a particle that displaces as a response to the flowing LC \citep{MusevicSoftMatter2017}. 
Thermophoresis of colloids in nematic LCs has also been reported \cite{kolacz2020thermophoresis}, although induced by an externally applied temperature gradient that spans 
the whole LC sample. That is different from self-thermophoresis, where the 
temperature gradient is local, self-generated, and occurs at a microscopic scale 
(\textit{i.e.} along the microparticle). We are not aware of previous reports of particles driven by self-thermophoresis swimming in nematic LCs. 

\textcolor{black}{It is worth mentioning the extensive recent experimental 
\cite{ZhouPNAS2014,mushenheim2014dynamic,turiv2020polar,
zhou2017dynamic} and theoretical 
\cite{genkin2017topological,lintuvuori2017hydrodynamics,chi2020surface,
soni2018enhancement,krieger2015microscale} 
work that has focused on the study of living swimming 
bacteria in lyotropic LCs. The complex dynamic phenomena that emerge 
from the coupling between the bacteria-generated flow field 
and the long-range orientational order of the lyotropic LCs have been studied in detail. 
Many of the phenomena observed and 
studied in the aforementioned systems could be relevant to the motion 
of synthetic active particles within thermotropic LCs. However, the details 
of the swimming mechanism of self-thermophoretic particles in thermotropic LCs 
can be significantly different from the swimming mechanism of bacteria
inside lyotropic LCs. In general, active particles within thermotropic LCs have the 
potential of exhibiting similar complex phenomena to the biological counterparts.
Moreover, the purely synthetic systems  could be easier to prepare experimentally,
making the study of these complex systems simpler and more accessible.
The aforementioned simplicity of the purely synthetic systems can in turn
increase the scope of their technological applicability.}

In this paper, we report the self-propulsion of light-activated Janus particles coated with a light-absorbent titanium layer in a thermotropic nematic LC. During illumination, the coated side of the particles self-induces a local nematic-isotropic phase transition to self-propel perpendicular to the nematic director and towards the self-generated isotropic region. The particle trajectories were tracked at different intensities of the applied light. We developed a phenomenological model to describe the MSD of the Janus particles. This model is fitted to the experimental data to test several hypotheses about the swimming mechanism and dynamics of the particles in the LC. The model suggests the possibility of utilizing Janus particles as probes for measuring the 
rheology of nematic LCs. 
The relatively simple system proposed here can serve as a biomimetic 
surrogate for studying transport mechanisms in complex biomaterials.
Previous work on  the motion of colloids embbeded in LCs has mostly focused on motion 
in the direction parallel to the nematic director. Here we focus on the active motion orthogonal to the orientation of the mesogens.  Our main goal is to explore the dynamics of individual colloids influenced by a highly structured non-isotropic medium and the collective motion triggered by the inter-colloidal elastic and surface interactions not observed in isotropic liquids. \textcolor{black}{This represents an important contribution to developing a platform for highly structured anisotropic active materials.}

\section{EXPERIMENTS}\label{experimental}

\begin{figure}[h t]
\center
\begin{overpic}[width=\linewidth]{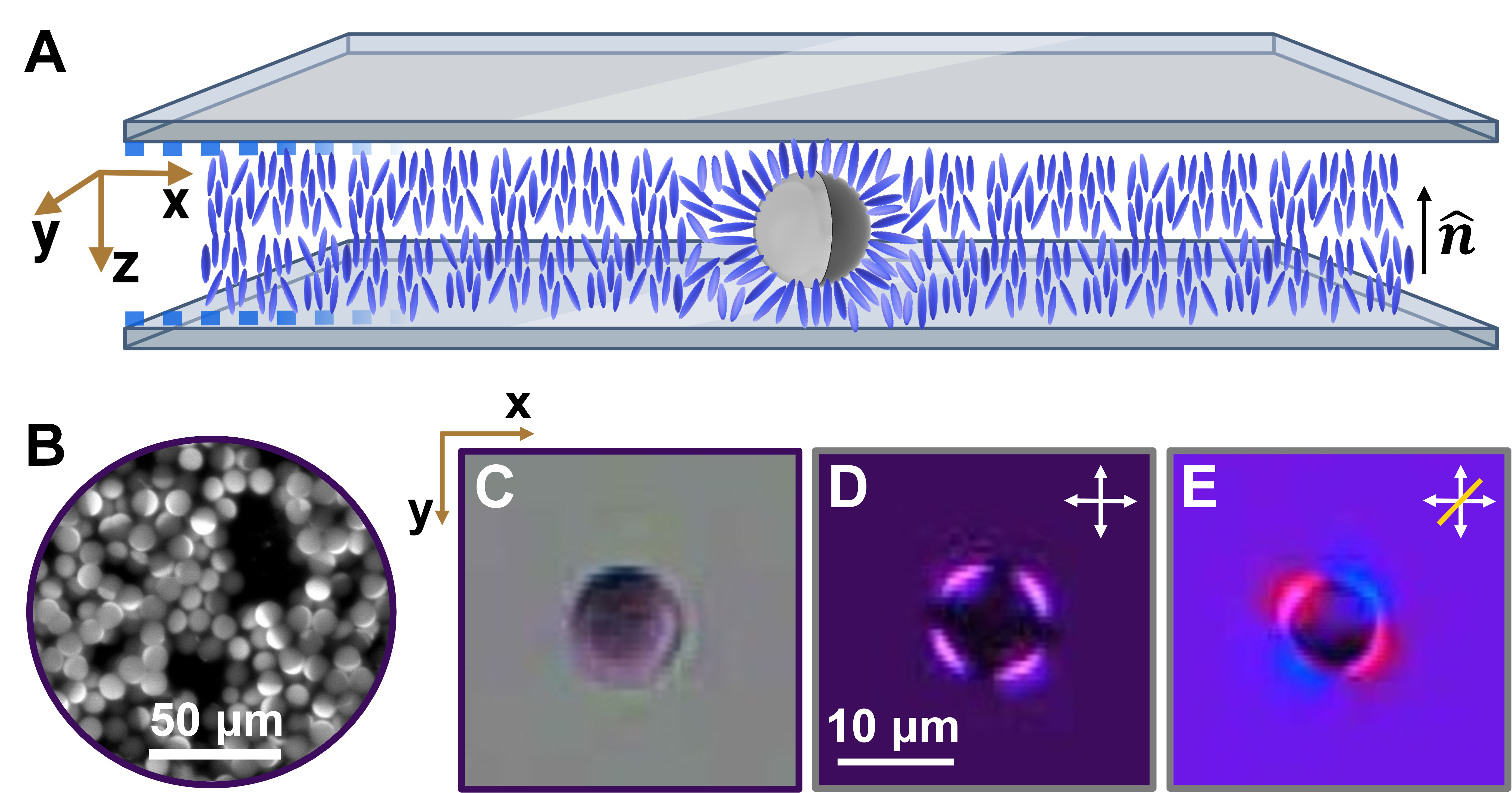}
\end{overpic}
\caption{Passive Janus particles. A) Non-scaled sketch of a Janus particle confined within a 5CB-glass cell. The unit vector $\hat{\bm{n}}$ is the LC nematic director. B) Scanning Electron Microscopy image. The bright sections correspond to the Ti-coated side of the particles. C) Optical microscope bright-field picture of a Janus particle. D) Cross-polarized picture of a Janus particle. The birefringent bright lobes are characteristic of the homeotropic LC anchoring on the colloid's surface. E) Particle imaged using a full-wave retardation plate. The red and blue colors corroborate the orientation of the mesogens perpendicular to the particle.}
\label{passive}
\end{figure}

\begin{figure*}[t]
\centering
\includegraphics[width=0.8\linewidth]{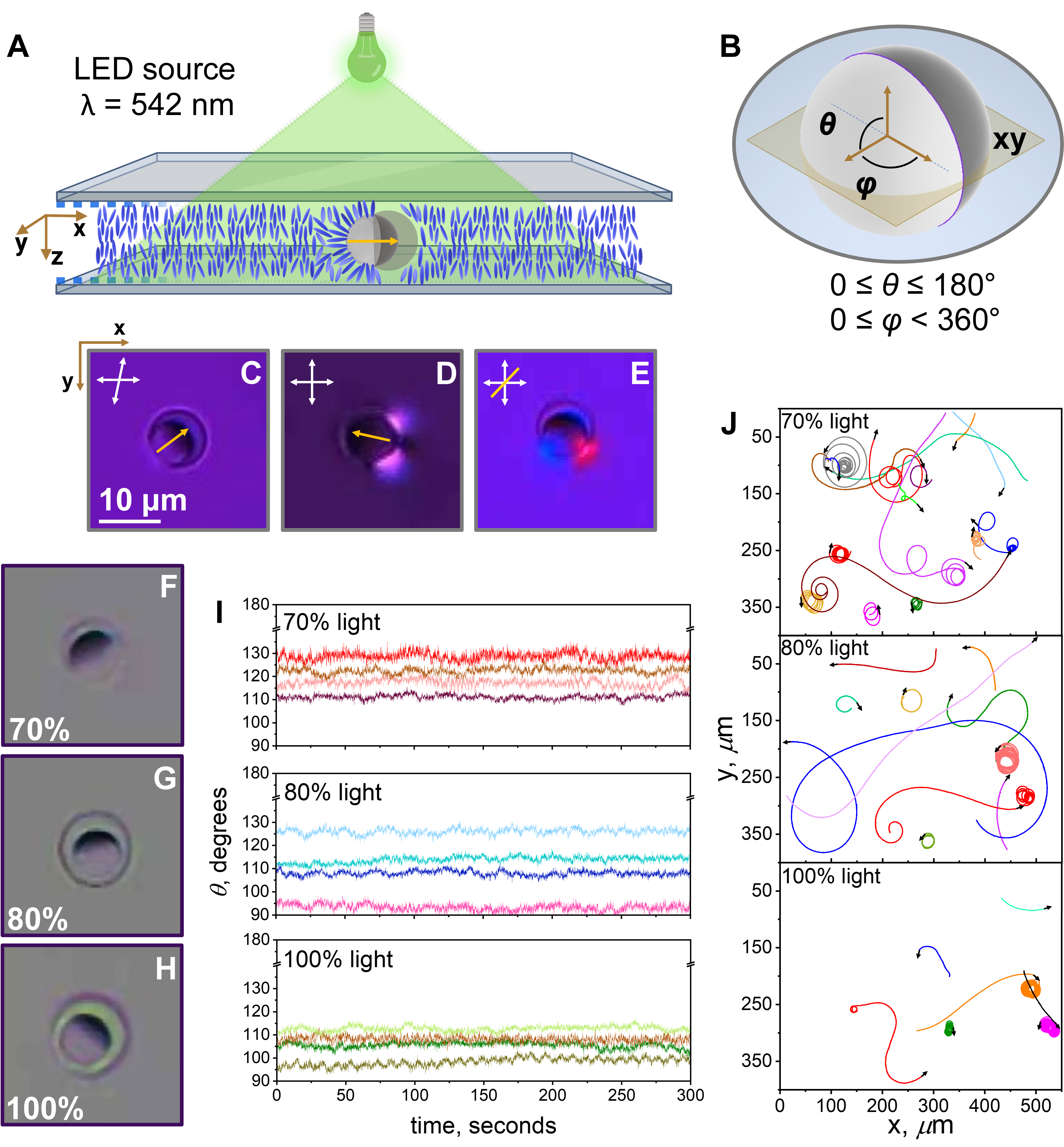}
\caption{Active Janus particles. Visualization and computation of orientations and trajectories. A) Sketch of a Janus particle activated using an LED source. The colloid always displaces perpendicular to the nematic director. The shadowed area represents the self-induced isotropic phase. B) \update{Sketch of the in-plane and out-of-plane orientations of a Janus particle. The $0\degree \leq \theta \leq 180\degree$ angle accounts for out-of-plane rotations. The minimum value corresponds to a coated side fully oriented upwards. The $0\degree \leq \varphi < 360\degree$ angle accounts for in-plane rotations.} C) Slightly rotated polarized image. The particle moves towards the direction of the isotropic phase. D) Cross-polarized picture of an activated Janus particle. The rear lobes are preserved. The yellow arrows indicate the motility direction. E) Corresponding image with a full-wave retardation plate. The colors indicate the preservation of the anchoring on the cooler side of the particle. F), G), H) Bright-field microscopy images at 70$\%$, 80$\%$, and 100$\%$ light power illumination, respectively. The area covered by the isotropic phase increases with the light intensity. I) Graphs of the $\theta$ angle for the different light intensities experiments over 5 min. Each graph shows four plots for different particles. In all cases, the angle is essentially constant on average. J) In-plane trajectories followed by different Janus particles for all the applied light intensities and for different experiments. Overlaps do not represent actual overlapped paths from experiments. The trajectories do not reflect a particular preferred orientation. The arrows point in the direction followed by each particle.}
\label{active}
\end{figure*}

Janus particles are prepared by standard methods \cite{musevicOptExpress2010} using $10$ $\mu$m silica spheres functionalized with amine groups (CD Bioparticles, NY USA). A 30 nm layer of titanium was deposited on top of the particles (AJA ATC-Orion 8E e-beam evaporation system, AJA International Inc, MA USA) (See Fig.\ref{passive}B for SEM verification; FEI Quanta 650 FEG SEM, Thermo Fisher Scientific Inc, USA). Afterward, to induce LC surface homeotropic anchoring on the Janus particles, they are suspended in a $2$ $\%$ deionized water solution ($18.2$ M$\Omega$ cm, Millipore, Bedford, MA USA) of dimethyloctadecyl[3-(trimethoxysilyl)propyl]ammnonium chloride (DMOAP) (Sigma-Aldrich, USA) for 10 min, then rinsed profusely with deionized water, and finally baked at 100$^\circ$C for one hour. It is well known that DMOAP chemically functionalizes silica particles after a hydrolysis reaction of the silane molecules that bond to the oxygen-silica molecules. The high-temperature curing process permits the polymerization of the silane monomers to form polysiloxanes. However, in our case, amine groups (NH$_2$) do not disturb silanization. On the contrary, it has been shown that these groups may improve the chemical reactions for various silane molecules \cite{polymers2018}. The capped side of the Janus particles is also silanized, corroborated by previous tests done on TiO$_2$ surfaces \cite{RSCAdvances2020}. The dried Janus particles are mixed at a dilute concentration of 0.01 wt $\%$ in 4-Cyano-4'-pentylbiphenyl (5CB) LC, purchased from Hebei Maison Chemical Co., Ltd (China). Thermogravimetric measurements were done with a TGA instrument (TA Instruments, USA) to ensure final particle concentration. We build cells to confine the LC-particle suspension for experimental observations using soda-lime glass microscope slides (Thermo Scientific, NH USA), cleaned with deionized water and air plasma treatment to eliminate any trace of organic pollutants. Later, the glass slides are also treated with DMOAP. 12 $\mu$m thick mylar films are implemented as cell spacers (Premier Lab Supply, FL USA), and 5-minute epoxy resin is used to seal the cells (Devcon, MA USA). \update{Slight variations in the cell thickness are possible owing to the method of assembling.} The LC-colloidal suspension is then heated to 50$^\circ$C and gently introduced into the glass cells by capillary effects. The cells are immediately sealed with 5-minute epoxy resin and taken apart from the hot plate to quench the system; otherwise, the Janus particles would reorganize and cluster since they are prone to be dragged by the nematic-isotropic (NI) interface. The LC is allowed to relax to get perfect homeotropic alignment. We use the AURA III LED engine (Lumencor, OR USA) as the activation source (bandpass filter $\lambda$ = 542 nm, FWHM = 33 nm. Maximum nominal output power = 500 mW). Light-power measurements are done using a Thorlabs PM160T power meter (USA). The light source is assembled into a Leica DM-2700P microscope (Germany) in reflection mode using a liquid light guide and a collimator. We use a custom-made filter cube (Chroma Technologies, USA; dichroic mirror $\%T \geq 95\%$ for 470 nm $\leq \lambda \leq$ 480 nm, $\%R \geq 95\%$ for 495 nm $\leq \lambda \leq$ 545 nm, and transmission filter $\%T \geq 95\%$ for 445 nm $\leq \lambda \leq$ 470 nm and 605 nm $\leq \lambda \leq$ 650 nm). The collimated light beam passes through an N-Plan achromatic objective ($20\times$). Simultaneously, a white LED source is used in transmission mode to image the sample. Bright-field and polarized movies are recorded at 15.3735 fps with a Leica MC170 HD camera. The camera's spatial resolution is 0.346620 $\mu$m/pixel. Therefore the minimum MSD, $\langle \Delta r_{{\rm b},\perp}^2(t)\rangle$, that can be measured is of the order of $\sim 10^{-2} ~\mu\text{m}^2$ and the minimum  measurable \update{lag-time} is $\sim 0.065$ s (\textit{i.e.} sampling frequency of 15.4 Hz). \update{We use a dual top-bottom heating stage HCS402 and a high precision temperature controller mk2000 (resolution 0.001$^\circ$C; INSTEC, USA) to maintain samples at constant $35.40 ~\pm 0.01^\circ$C \textcolor{black}{in the nematic phase}. Fig \ref{passive}A shows a schematic of a Janus particle within the LC-glass cell (non-scaled).} Fig \ref{passive}C presents a bright field picture of a passive Janus particle, and Fig \ref{passive}D shows an example of a particle in between crossed polarizers. A well-defined birefringence is observed, with four bright lobes characteristic of the LC homeotropic anchoring on the particle’s surface, verified using a full-wave retardation plate (see Fig \ref{passive}E). The central dark cross suggests the creation of a Saturn ring formed at the center of the colloid, located on a plane parallel to the observation plane of view. \update{To activate the particles, the Aura engine illuminates the whole field of view of the microscope. The energy and nature of the light source prevent the LC from being directly altered; therefore, no polarization of the light is required.} The coated side of the particles absorbs the light and consequently heats up, inducing a local NI transition of the LC. This symmetry-breaking triggers the motility of the colloids (see the Supporting Information SI, MovieS1). Fig \ref{active}A shows a schematic of a transversal view of the particle’s activation. Note that the particle always moves perpendicular to the LC director vector $\hat{\bm{n}}$ \update{and in the direction of the self-generated isotropic phase. Moreover, the particle remains passive before forming the local isotropic phase, even when illuminated with the green light. This suggests that the symmetry breaking 
that generates the active transport occurs when the surface interactions on one side of the particle are released, and the elastic energy locally decreases near the isotropic phase.} Experimentally, the Janus colloid is observed as in Fig \ref{active}C forming a local isotropic phase when illuminated with constant and homogeneous light (non-perpendicular polarizers), or as in Fig \ref{active}D in between crossed polarizers, where the rear birefringence lobes are preserved. Fig \ref{active}E is the corresponding observation with the full-wave retardation plate. The yellow arrows indicate the direction of the particle's motility (defined as vector $\bm{u}_{\rm b}$). Experiments were carried out at three different light intensities (70$\%$, 80$\%$, and 100$\%$), described as a percentage value out of a maximum light power of $106.90~\pm 0.01$ mW. Figs \ref{active}F-H show the corresponding particles (bright-field) for the different light intensities. We emphasize that the LC does not absorb light in this region of the electromagnetic spectrum. Therefore, the effect of temperature increase always occurs locally at the particle level.

Bright-field movies were used to facilitate the tracking analysis, and most of the movie processing is done with the ImageJ Fiji distribution software \cite{Fiji}. To properly track the particle center of mass and reconstruct the particle trajectories, we follow the protocols found elsewhere \citep{CUI2017452, Granick}. In brief, standard methods track the brightest spot found on an image, such as the center of plain colloids. However, when using Janus particles, those identified spots are off-center. To translate the tracked coordinates to the center of mass, a computation of the out-of-plane angle of rotation ($\theta$) and the in-plane angle ($\varphi$) is necessary (see Fig \ref{active}B). Calculating $\theta$ requires acquiring grayscale images and comparing the gray mean value for an active Janus particle with the value of a Janus particle with $\theta = 0$, which is a particle with the fully coated side facing up. \textcolor{black}{Consequently, with this convention, $\theta = 180\degree$ corresponds to a Janus particle with the uncoated side facing up.} Values of the acquired angles can be seen in Fig \ref{active}I for each different applied light intensity. Each graph shows the variations of $\theta$ during 5-minute experiments for four different particles. In all cases, the angles essentially do not change, only within certain \update{fluctuations attributed to the image analysis at the pixel level}. Also, we found that for the active Janus particles embedded in the nematic LC, the angles vary from $\sim 95\degree$ to $\sim 130\degree$. To compute the in-plane $\varphi$ angle, the "analyze particle" module of the Fiji software is used. This permits the fit of an averaged ellipse around the coated side of the particles, with the $\varphi$ angle and coordinates of the ellipse's center as output values. The full trajectories and variations of $\varphi$ of different active Janus particles during the 5-minute experiments are shown in Fig \ref{active}J for the different cases studied in this work (see examples in the SI. MovieS2, MovieS3, and MovieS4 for intensities of 70$\%$, 80$\%$, and 100$\%$ of the applied light respectively). Notice the diverse pathways the particles \update{undergo}, indicated by black arrows at the end of each trajectory. The overlaps of the pathways do not represent real overlaps in the experiments and are plotted together for comparison purposes. The pathways are displayed using Trackpy \cite{Trackpy} using the center of mass particle coordinates.

\section{THEORY}\label{theory}

Here, we employ a generalized Langevin equation (GLE) to describe the motion of the 
Janus particle in the nematic LC,
\begin{align}\label{1_GLE}
m \frac{d^2 \bm{r}_{\rm b}(t)}{dt^2}&=
\bm{f}_{\rm E}(t)-\int_0^{\infty}\bm{\zeta}(t-t')\cdot\frac{d \bm{r}_{\rm b}(t')}{dt'} dt'+
\bm{f}_{\rm B}(t),
\end{align}
where $m$ is the particle mass, $\bm{r}_{\rm b}(t)$ is the particle position,
$\bm{\zeta}(t)$ is the time-dependent friction tensor, 
$\bm{f}_{\rm E}(t)$ is the active force and
$\bm{f}_{\rm B}$ are Brownian forces that obey the fluctuation-dissipation theorem (FDT),
\begin{align}
\left\langle \bm{f}_{\rm B}(t)\bm{f}_{\rm B}(t')\right\rangle_{\rm eq}
=k_{\rm B}T \bm{\zeta}(t-t'),
\end{align}
where $k_{\rm B}$ is the Boltzmann constant and $T$ is the temperature.

We will assume that $\bm{f}_{\rm E}:=A\dfrac{\bm{u}_{\rm b}(t)}{\ell_0}$ 
where $A$ is an unknown function of the intensity 
of the applied light, $\ell_0$ is the diameter of the Janus particle
and $\bm{u}_{\rm b}(t)/\ell_0$ is its orientation vector. $\bm{u}_{\rm b}$ points 
from the bare side of the Janus particle to 
the metal-coated side of the particle. \textcolor{black}{We assume that
$t=0$ is the time at which the self-generated local isotropic phase 
around the metal-coated side of the Janus particle has reached its
steady-state size. As shown in Figs \ref{active}F--H 
the size of the self-induced isotropic phase
around the metal-coated side of the Janus particle 
increases with increasing the intensity of the applied light. 
Therefore it is expected 
that the parameter $A$ will also exhibit some dependence on the 
intensity of the applied light.}

To model the rotation of the Janus microparticle, we
treat it as a dumbbell of rest length $\ell_0$. 
The overdamped dynamics for the end-to-end vector of a Janus dumbbell 
can be written as \cite{cordoba2020simple},
\begin{align}\label{ecQ1}
\frac{d \bm{u}_{\rm b}(t)}{dt}=-\frac{1}{\lambda_{\rm r}} \left( 1-\frac{\ell_0}{u_{\rm b}}\right) 
\bm{u}_{\rm b}(t)+\bm{g}_{\rm B}(t).
\end{align}
Here $u_{\rm b}=\sqrt{\bm{u}_{\rm b}\cdot\bm{u}_{\rm b}}$ is the magnitude of $\bm{u}_{\rm b}$, $\lambda_{\rm r}$ 
is a constant with units of time 
and the Brownian velocities, $\bm{g}_{\rm B}$, obey the FDT,
\begin{align}\label{fdtg}
\left\langle \bm{g}_{\rm B}(t)\bm{g}_{\rm B}(t')\right\rangle_{\rm eq}
=\frac{k_{\rm B}T}{\zeta_{\rm r}} \delta (t-t')\bm{\delta},
\end{align}
where $\zeta_{\rm r}$ is a friction coefficient.

To account for both the anisotropic viscous response of the nematic LC and
the viscoelasticity  arising from the creation and destruction of defects 
in the nematic field as the particle moves, we use
the following form for the friction tensor,
\begin{align}\label{zeta}
\bm{\zeta}(t)=\delta(t) \bm{\zeta}_0+H(t) \bm{\delta} K e^{-t/\lambda}.
\end{align}
Where $\delta(t)$ is the Dirac delta function, $H(t)$ is the Heavy side step function,
$K$ and  $\lambda$ are the strength and relaxation time of the viscoelastic element
respectively and $\bm{\delta}$ is the identity tensor.
For a spherical particle suspended in a nematic LC, the purely viscous part of 
the friction tensor in eq. (\ref{1_GLE}) is given by,
\begin{eqnarray}
\bm{\zeta}_0=
\left(\begin{array}{cc}
\zeta_{0,\parallel} & 0\\
0 & \zeta_{0,\perp}
\end{array}\right).
\end{eqnarray}\label{comp}
$\parallel$ represents the direction parallel to the director vector, $\hat{\bm{n}}$, and 
$\perp$ is the direction perpendicular to $\hat{\bm{n}}$. 
$\zeta_{0,\parallel}$ and $\zeta_{0,\perp}$ are \cite{cordoba2016anisotropy,gomez2013flow},
\begin{eqnarray}\label{zeta0}
\zeta_{0,\parallel}&=&\frac{4\pi R(\eta_c-\eta_b)}
{\frac{\eta_c}{\eta_b}\frac{{\rm arctan} \left(\sqrt{\eta_c/\eta_b-1}\right)}
{\sqrt{\eta_c/\eta_b-1}}-1}, \\
\nonumber
\zeta_{0,\perp}&=&\frac{8\pi R(\eta_c-\eta_b)}
{1-\frac{{\rm arctan} \left(\sqrt{\eta_c/\eta_b-1}\right)}
{\sqrt{\eta_c/\eta_b-1}}+
\frac{\eta_c-\eta_b}{\eta_a}\frac{{\rm arctan} \left(\sqrt{\eta_c/\eta_a-1}\right)}
{\sqrt{\eta_c/\eta_a-1}}}.
\end{eqnarray}
Here $R=\ell_0/2$ is the radius of the particle, $\eta_a=\alpha_4/2$,
$\eta_b=(\alpha_3+\alpha_4+\alpha_6)/2$ and
$\eta_c=(-\alpha_2+\alpha_4+\alpha_5)/2$. 
$\{\alpha_i\}, i=1...6$ are the six
Leslie viscosity coefficients. According to Parodi's relation
$\alpha_6=\alpha_2+\alpha_3+\alpha_5$.

Note that the expressions given in eq. (\ref{zeta0}) were derived under the assumption that there are no director field gradients in the LC.
This leads to a no-torque, boundary condition
that is imposed by default, which implies insignificant surface
anchoring energy \citep{rey2002dynamical,gomez2013flow}.
With the assumptions described above, the friction tensor that is obtained
for a particle suspended in a LC is purely viscous,
and accounting for the viscoelasticity of the medium is no longer possible.
In the linear response regime, however, the
correspondence between the creeping flow equations of motion for a
purely viscous fluid and for a viscoelastic material in the frequency domain
\citep{Lee1955, Zwanzig1970, Xu2007,cordoba2016anisotropy} 
can be exploited to obtain the response function for an anisotropic viscoelastic material.
The specific form of the friction tensor used in eq. (\ref{zeta}) is inspired by 
the correspondence principle. Therefore eq. (\ref{zeta}) 
remains valid as long as $\bm{f}_{\rm E}$ is small enough so that the 
fluid remains within the linear viscoelastic regime.

The use of eq. (\ref{zeta0}) also assumes that the Janus 
particle feels the medium in which is embedded as a continuum
with an anisotropic viscosity equal to that of the bulk nematic LC. We 
do not consider here the details at the particle-medium 
interface that lead to the emergence of the swimming force. 
In the model proposed, those details are implicitly accounted
through $\bm{f}_{\rm E}$. Moreover eq. (\ref{zeta}) implies that 
the Janus particle does not interact either through hydrodynamic interactions
and/or elastic forces with other Janus particles. That is a safe assumption for the analysis
performed in this work. \update{Ongoing experimental 
and theoretical work focuses on the details of the swimming mechanism and on the dynamics of interacting Janus particles in a nematic LC.}

Eqs. (\ref {1_GLE}) and (\ref{ecQ1}) are solved in the Laplace domain to derive an expression for 
the MSD perpendicular to $\hat{\bm{n}}$, $\left\langle \Delta r_{{\rm b},\perp}(t)^2\right\rangle$,
of the Janus particles embedded in a nematic LC,  
\begin{align}\label{MSDfull}
\langle \Delta 
r_{{\rm b},\perp}^2(t)\rangle
=&\frac{k_{\rm B}T \left[ \left( 1- e^{-\Theta t} \right)
K \lambda^2+t \Theta \lambda \zeta_{0,\perp}
\right]}{\left(\Theta \lambda \zeta_{0,\perp}\right)^2} \\ \nonumber
+&\frac{A^2 e^{-2\Theta t}\left[K \lambda^2 - e^{\Theta t}\left( K \lambda^2+
t\Theta \lambda \zeta_{0,\perp}\right)\right]^2}{(\Theta \lambda \zeta_{0,\perp})^4}
\\ \nonumber
+&\frac{k_{\rm B} T \lambda_{\rm r}^2 A^2}{2\ell_0^2\zeta_{0,\perp}
\zeta_{\rm r}(\Theta \lambda \zeta_{0,\perp})^3} \bigg[ 
2 t \zeta_{0,\perp}^2 \Theta \lambda -K^2\lambda^3e^{-2\Theta t} \\ \nonumber
+&K\lambda^2(4\zeta_{0,\perp}+K\lambda)
-4K\zeta_{0,\perp}\lambda^2
e^{-\Theta t}\bigg],
\end{align}
where $\Theta:=\dfrac{K}{\zeta_{0,\perp}}+\dfrac{1}{\lambda}$.
This model has six adjustable parameters $A$, $\lambda_{\rm r}$, $\zeta_{\rm r}$
$\zeta_{0,\perp}$, $K$ and $\lambda$.
The details on how eq. (\ref{MSDfull}) is obtained are given in Appendix \ref{appA}.
Note that to obtain eq. (\ref{MSDfull}) we have assumed that $\bm{u}_{\rm b}$ is initially 
perpendicular to $\hat{\bm{n}}$
and does not significantly deviate from this direction during the course of the experiment, as has been observed (see Fig \ref{active}I).
We also assume that $u_{\rm b}$ does not significantly deviate from $\ell_0$.
Which means that eq. (\ref{MSDfull}) is valid for $\zeta_{\rm r}\gg\zeta_\perp$ and
$t\gg\lambda_{\rm r}$. When deriving eq. (\ref{MSDfull}) we also 
assume that the particle inertia is negligible (\textit{i.e.} $m\approx0$) 
at the time scales where the measurements are performed. The density of
a nematic LC such as 5CB \cite{oweimreen1986density} 
at $35.4$ $^\circ$C is $1011~\text{Kg}/\text{m}^3$. 
For a $10~\mu$m particle that is buoyant in 5CB at that temperature, the 
inertial time scale is $m/\zeta_{0,\perp}\approx2\times10^{-7}$ s.
The shortest times that can be measured with the particle tracking 
employed in this work are $\sim6.5\times10^{-2}$ s. Therefore, neglecting particle inertia is a valid assumption in the analysis performed here.

By taking the limit $K\longrightarrow0$ in eq. (\ref{MSDfull})
a purely viscous model can be recovered,
\begin{align}\label{MSDvis}
\langle \Delta 
r_{{\rm b},\perp}^2(t)\rangle=
k_{\rm B}T\left(\frac{1}{\zeta_{0,\perp}} +
 \frac{A^2\lambda_{\rm r}^2}{\ell_0^2\zeta_{\rm r}\zeta_{0,\perp}^2}\right) t+
\left(\frac{A^2}{\zeta_{0,\perp}^2}\right) t^2.
\end{align}
This model has four adjustable parameters $A$, $\lambda_{\rm r}$, $\zeta_{\rm r}$
$\zeta_{0,\perp}$. The model given by eq. (\ref{MSDvis}) does not account
for the elasticity generated by defects in the nematic field of the LC.
This equation has a term that scales linearly with time 
(\textit{i.e.} a diffusive term) and 
a term that scales with time squared (\textit{i.e.} a ballistic term).
Note that the prefactor for the diffusive term includes the diffusivity, 
$\mathcal{D}_{\perp}:=k_{\rm B}T/\zeta_{0,\perp}$, but also has a second term that includes 
the active force parameter $A$ and the parameters related to the rotation of 
the particle. The ballistic term depends on the active 
force parameter, $A$, and on $\zeta_{0,\perp}$.

\textcolor{black}{It should be emphasized that 
the analytic expressions for the mean-squared displacement of the 
particles, eqs. (\ref{MSDfull}) and (\ref{MSDvis}), 
are expected to be valid only when $\zeta_{\rm r}\gg\zeta_\perp$.
In other words, eqs. (\ref{MSDfull}) and (\ref{MSDvis}) are derived under the assumption that 
rotations away from the plane perpendicular to the nematic director 
(\textit{i.e.} out-of-plane rotations) are very small during the course of the experiments.
If one allows for rotations of arbitrary magnitude, analytic solutions of the model 
equations are no longer possible. Moreover, the aforementioned assumption 
has been confirmed to be valid by measuring the out-of-plain $\theta$ angle.
This angle is shown in Fig \ref{active}I, where it can be observed that it does not significantly 
change during the course of the experiments. With respect to in-plane rotations, 
we estimate that the trajectories have in-plane persistence lengths ranging from $10$ to $30$ $\mu$m.
Therefore the effect of in-plane rotations of the particle
should not be relevant for mean-squared displacements, 
$\langle \Delta r^2_{\perp} (\tau) \rangle$, smaller than about
$10^2$ $\mu \text{m}^2$. Accordingly eqs. (\ref{MSDfull}) and (\ref{MSDvis}) 
should be valid up to those values of the mean-squared displacement. For larger displacements, in-plane rotations will become relevant, and a diffusive region will appear 
in the $\langle \Delta r^2_{\perp} (\tau) \rangle$ at long times.}

\section{RESULTS AND DISCUSSION}\label{results}

\begin{figure*}[t]
\centering
\includegraphics[width=\linewidth]{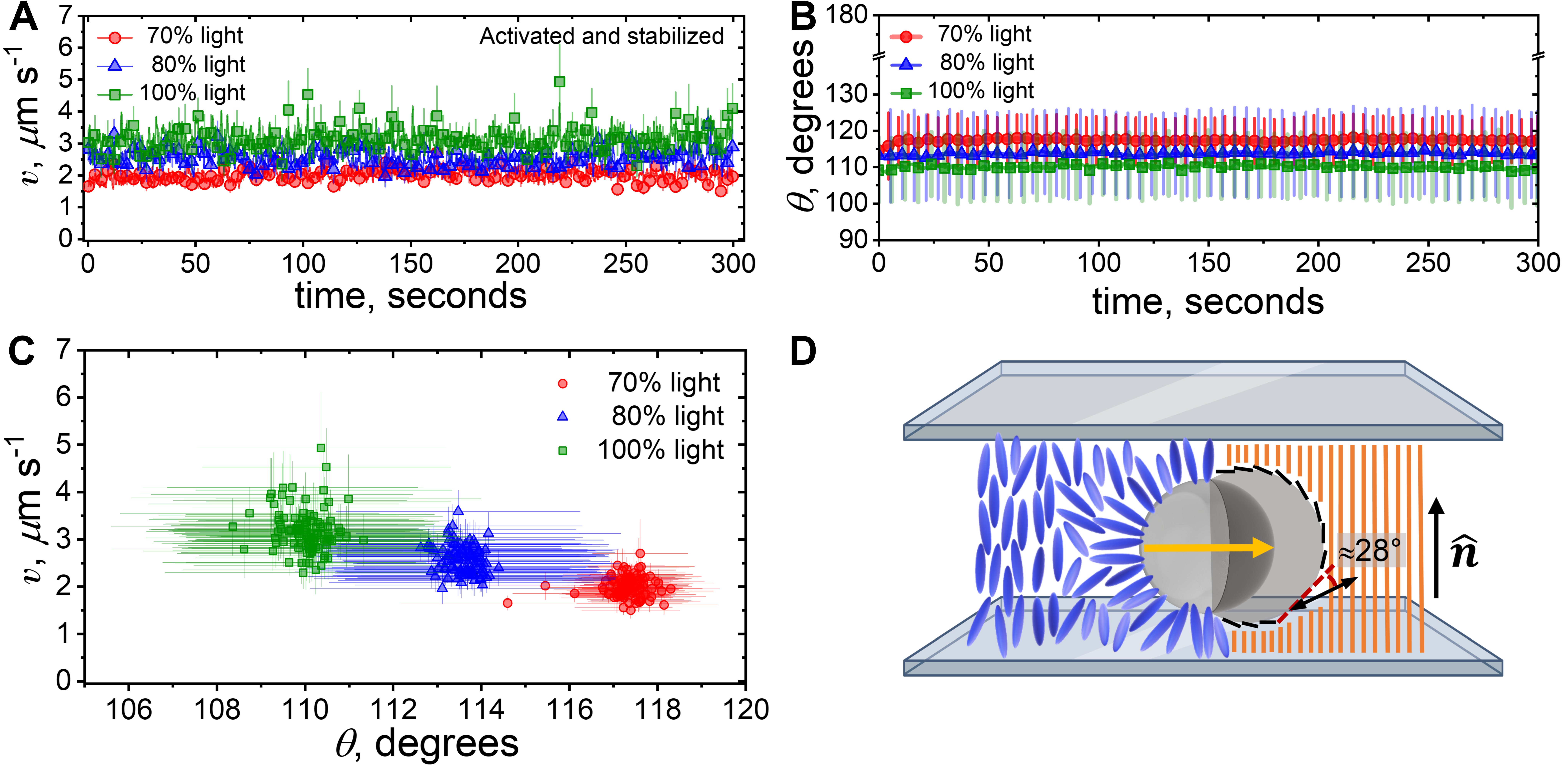}
\caption{\textcolor{black}{Instantaneous speeds of the active Janus particles, average $\theta$ angle, and sketch of particle stability. A) Instantaneous speeds captured for activated Janus particles. The active particles were allowed to stabilize for two minutes after illumination before recording the speeds. The data was recorded for a duration of 5 min. The average constant values clearly show increased speeds with increasing light intensity. The bars account for the standard error. B) Average values of the $\theta$ angle (\text{i.e.} out-of-plane rotations) over all particles analyzed. The reported time frame corresponds to the same considered in A). The $\theta$ values remain constant and show a dependence on the activation-light intensity. The bars represent the standard deviation. C) Relation of the particles' instantaneous speeds and the out-of-plane rotational angle. Smaller angles, corresponding to higher light intensities, increase the probability of reaching higher speeds. The average values are 3 $\mu$m/s, 2.5 $\mu$m/s, and 2 $\mu$m/s, for 100$\%$, 80$\%$ and 70$\%$ of the applied light respectively. The bars represent the standard error. D) 2D sketch of the transversal section of an active Janus particle confined in an LC glass cell (not drawn at scale). The LC homeotropic anchoring on the particle surface is preserved on the left side. On the right side, the self-induced isotropic phase is shown. The mesogens reorient at the NI interface at a value of $\approx 28\degree$ from the interface tangent line. An exaggerated angle representation is shown with a red tangential line and a double-sided black arrow. Shorter black lines around the NI interface show the alignment incompatibility with the bulk nematic director shown with vertical orange lines.}}
\label{speed_traj}
\end{figure*}

Several particles were tested at the different light intensities already mentioned. \textcolor{black}{To capture the effect on the particle’s kinetics, the instantaneous speeds of the particles 
were computed, $v = \sqrt{v_{x}^{2} + v_{y}^{2}}$, using data spanning 
5 min. To achieve systematic data collection, the experiments were recorded after 2 min of activation to ensure good stabilization of the Janus particles with the LC relaxed around the particles. Fig \ref{speed_traj}A shows the instantaneous speeds averaged over all the tested particles. Particles maintain a constant terminal speed over time, in agreement with active particle systems and models in which drag forces balance active forces. Although the values of the speeds overlap within the statistical uncertainty for all cases, the averaged values lie around 3 $\mu$m/s, 2.5 $\mu$m/s, and 2 $\mu$m/s for 100$\%$, 80$\%$ and 70$\%$ intensity of the applied light respectively. The average $\theta$ during the duration of the experiment was also calculated for all the particles tested at different light intensities (see Fig \ref{speed_traj}B). In all cases, $\theta$ remains constant, as was mentioned in the Experiments section. The $\theta$ values for all intensities of the applied light lie within the same range of the standard deviation. However, the trend shows that particles activated with higher light intensities develop a smaller out-of-plane angle. Fig \ref{speed_traj}C shows the relation of the averaged instantaneous speeds and the $\theta$ angle for all the tested particles. The plot clarifies that the probability of achieving higher speeds increases for smaller angles. With higher light intensities, the
size of the self-induced isotropic phase size increases, allowing the particles to stabilize at a range of smaller angles. This means that the probability of observing the coated side of the particle facing the camera increases with light intensity (see Fig \ref{active}B). This feature is an effect of the variability on the stability of the Janus particles that can be rationalized in the following manner. Given the incompatibility of the nematic director orientation in bulk and at the boundaries, with the orientation of the mesogens attached to the colloid surface, colloidal particles can levitate in an LC cell \cite{LavrentovichPRL2007}. In our case, the mesogens adopt a known orientation at the NI interface, with the major axis rotated approximately $28\degree$ from the tangent line of the interface \cite{InterfaceAngle}. This reorientation of the mesogens occurs along the complete interface and in 3D, making them highly incompatible with the homeotropic nematic director in bulk induced by the glass substrates. Meanwhile, the non-coated side of the particle remains with the LC homeotropic anchoring (see sketch in Fig \ref{speed_traj}D). This scenario explains the stability of the 
particle's orientation during the course of the experiments, 
with no significant changes of the $\theta$ angle. 
However, the particles stabilize at different $\theta$ angles for different light intensities.
This may be attributed to variations in the size of the interface which together with the incompatibility 
of the mesogen orientations around the particle influence 
the interactions of the Janus particles with the LC bulk. In general, when 
a larger degree of broken symmetry is produced around the particle a higher 
particle speed is observed.}

To obtain the MSD from the trajectories of Janus particles, the $\gamma$-position ($\gamma=x,y$) 
of the particle as a function of time, $r_{b,\gamma}(t)$, is tracked.
Where $t=i \Delta t$, $1/\Delta t$ is the sampling frequency of the 
particle tracking technique, $N$ is the total number of measurements and $i=0,1,2,3...,N$. The MSD is then computed as,
$\langle \Delta r^2_\perp (\tau) \rangle:=
\frac{1}{n}\sum_{\gamma}\sum_{i=1}^n \Delta r^2_{i,\gamma}(\tau)$
where $\Delta r^2_{i,\gamma}(\tau):=[r_{b,\gamma}(t+\tau)-r_{b,\gamma}(t)]^2$,
$n = N-\tau/\Delta t$, $\tau$ is the \update{lag-time} and
the sub-index $\gamma:= \{x, y\}$ indicates the 
spatial direction of the measurement.
To calculate $\langle \Delta r^2_\perp (\tau) \rangle$ and 
estimate its statistical uncertainty, $\sigma(\langle \Delta r^2_\perp (\tau)\rangle)$,
here we employ MUnCH \cite{cordoba2022munch,MUnCH}. By using 
repeated block transformations, MUnCH can correctly estimate 
the statistical error of any autocorrelation at any given \update{lag-time}.
A common omission in the calculation of $\sigma(\langle \Delta r^2_\perp (\tau)\rangle)$
is to neglect the correlations inherent in the bead position data. 
These correlations are important in viscoelastic materials,
the uncertainty can be underestimated by a factor of 
up to $20$ if the correlation in the bead position data is neglected \cite{cordoba2022munch}.

\begin{figure}[]
\center
\includegraphics[width=\linewidth]{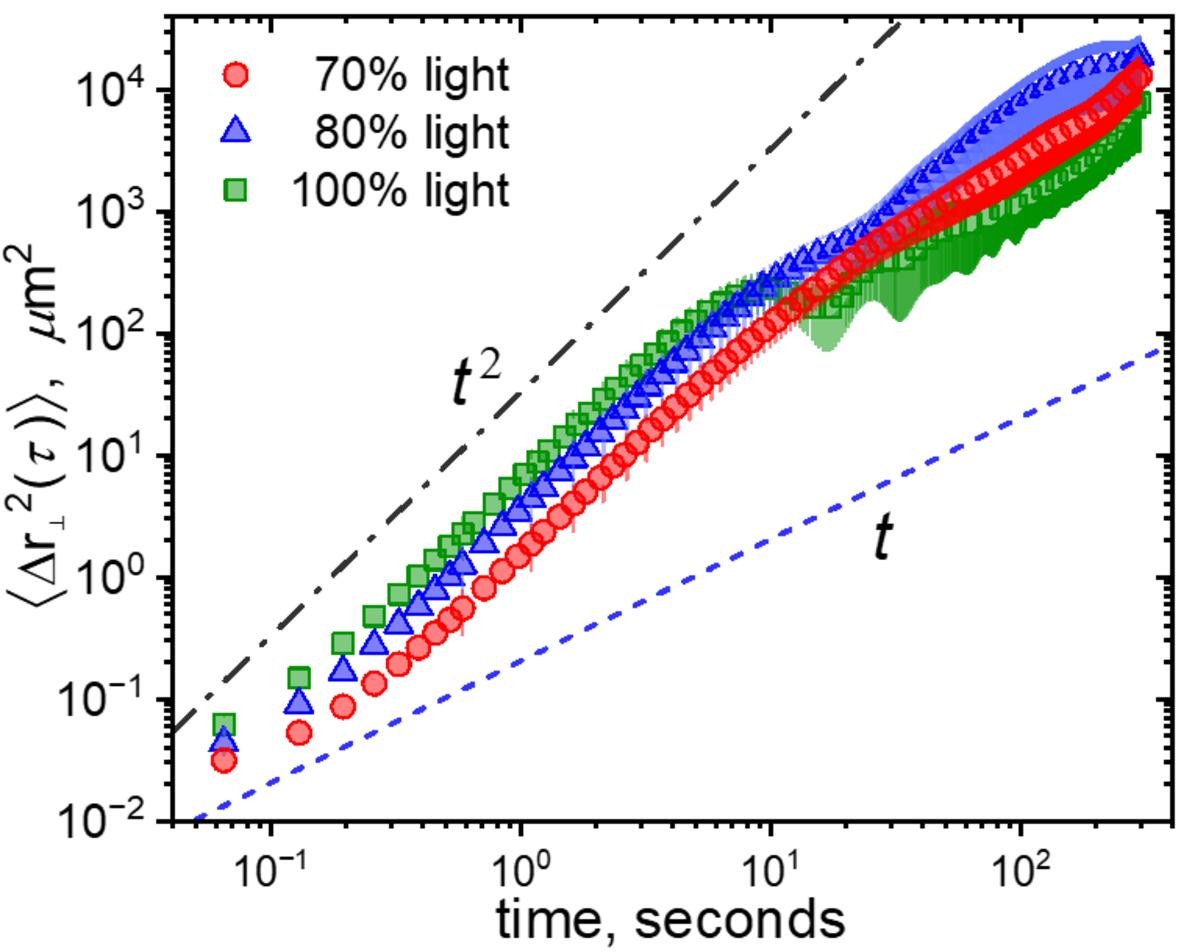}
\caption{\textcolor{black}{MSD computed from experimental trajectories for all the applied light intensities. The plots show a clear ballistic regime between $4\times 10^{-1}$ s and 10 s. At longer times, the shape of the curves captures the different pathways the particles take, either straight or circular.}}
\label{MSD}
\end{figure}

\begin{figure}[t]
\center
\includegraphics[width=\linewidth]{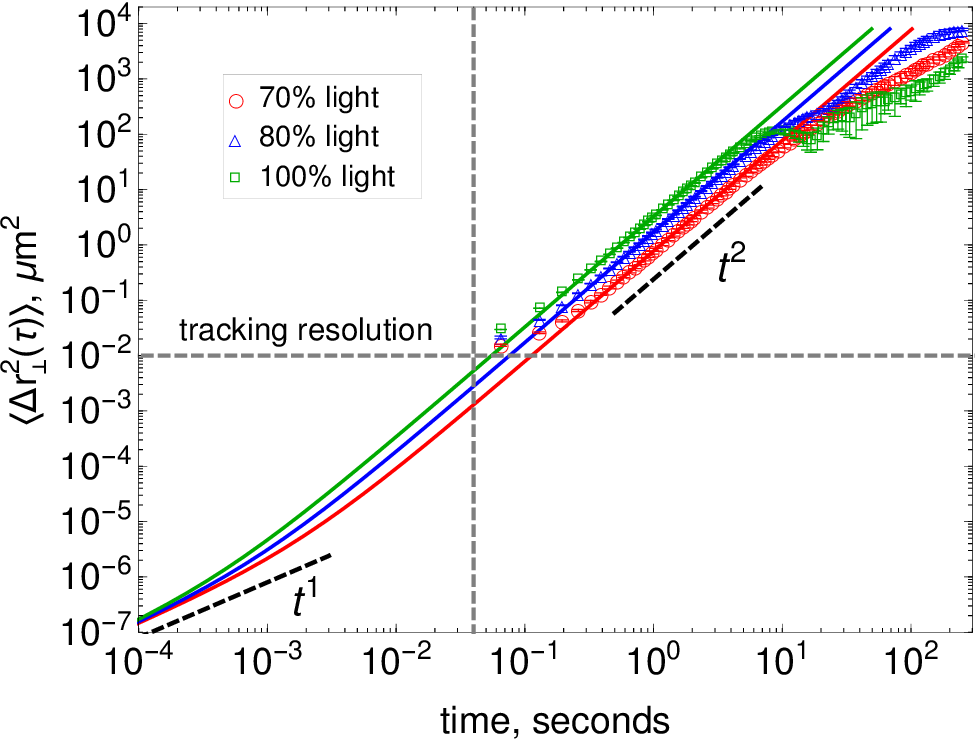}
\caption{Analysis of the mean-squared displacements calculated 
from the Janus particle trajectories 
using the purely viscous model. 
The symbols are calculated using the particle tracking data and 
the continuous lines are fits of eq. (\ref{MSDvis}) to the data. 
The value of the fitted parameter for each intensity of the applied light is,
$A=2.70 \times 10^{-12}~\text{N}$ (70\% intensity), $A=4.01 \times 10^{-12}~\text{N}$
(80\% intensity) and $A=5.55  \times 10^{-12}~\text{N}$ (100\% intensity).
Other parameters are set to $\ell_0=10~\mu\text{m}$, $\lambda_{\rm r}=1\mu\text{s}$,
$\zeta_{\rm r}=10 \zeta_{0,\perp}$.}
\label{fig1}
\end{figure}

Fig \ref{MSD} shows the ensemble averaged $\langle \Delta r^2_{\perp} (\tau) \rangle$ from the trajectories of several Janus particles embedded in the nematic LC. Note that a ballistic regime is found at intermediate lag-times. This ballistic regime 
coincides with the fairly straight paths that the particles follow during ten-second-long steps
of the experiments. At longer times, the ensemble averaged $\langle \Delta r^2_{\perp} (\tau) 
\rangle$ approaches a diffusive regime. At those long times, of hundreds of seconds, 
the paths taken by the particles include both straight and circular motions (see Fig \ref{active}J). 
We propose three possible contributions to the dynamics of the particles 
at those long times. A non-homogeneous temperature increase on the particle's surface since the light-absorbing titanium layer is not uniform. Another effect is the possible hydrodynamic interaction 
between particles that come close to each other. But
we expect that the more important contribution is the long-range inter-particle interactions driven by elastic distortions within the LC bulk. The particles deflect when encountering other particles at distances comparable to tens of the particle's diameter size. In some cases, the particles also deflect from their trajectories when swimming close to an impurity in the LC cell. 
The latter behavior, in particular, is characteristic of the highly structured non-isotropic medium
and is not expected to occur in isotropic liquids (see MovieS5 and MovieS6 in the SI to exemplify this behavior for intensities of 70$\%$ and 80$\%$ of the applied light). To fit eq. (\ref{MSDfull}) or eq. (\ref{MSDvis})
to this $\langle \Delta r^2_{\perp} (\tau) \rangle$ we first set $\zeta_{0,\perp}$ 
using the values of the Leslie viscosity coefficients reported \cite{chmielewski1986viscosity} 
for 5CB at $35.4~^\circ$C, $\alpha_2=-20 ~\text{mPa}\cdot\text{s}$,
$\alpha_3=0 ~\text{mPa}\cdot\text{s}$,
$\alpha_4=60 ~\text{mPa}\cdot\text{s}$,
$\alpha_5=0 ~\text{mPa}\cdot\text{s}$ and
$\alpha_6=-20 ~\text{mPa}\cdot\text{s}$. With these values, one obtains 
$\eta_{\rm a}=0.03~\text{Pa}\cdot\text{s}$, 
$\eta_{\rm b}=0.02~\text{Pa}\cdot\text{s}$ and $\eta_{\rm c}=0.04~\text{Pa}\cdot\text{s}$.
Using these values and $R=5~\mu\text{m}$ in eq. (\ref{zeta0}) one obtains
$\zeta_{0,\perp}=3.1\times10^{-6}~\text{N}\cdot\text{s}/\text{m}$ and
$\zeta_{0,\parallel}=2.2\times10^{-6}~\text{N}\cdot\text{s}/\text{m}$.
Other model parameters that are set when fitting eq. (\ref{MSDfull}) or eq. (\ref{MSDvis})
are $\ell_0$, $\lambda_{\rm r}$ and $\zeta_{\rm r}$. The diameter 
of the Janus particles, $\ell_0$, is known and equals $10~\mu$m. The 
parameter $\lambda_{\rm r}$ is set to a value much smaller than the time 
resolution of the experimental measurements. By doing so, the assumptions 
used when deriving eq. (\ref{MSDfull}) or eq. (\ref{MSDvis}) are satisfied, 
and the particle effectively behaves as a rigid body at the time scales of interest. 
The rotational friction coefficient of the particles is assumed to be 
$\zeta_{\rm r}=10 \zeta_{0,\perp}$. This assumption is based on the observation
that the Janus particles do not significantly rotate away from the focal plane 
(\textit{i.e.} the plane perpendicular to $\hat{\bm{n}}$) during the course of the 
experiments.

Fig \ref{fig1} shows the fits of the purely viscous model, eq. (\ref{MSDvis}), 
to the $\langle \Delta r^2_{\perp} (\tau) \rangle$ calculated from the particle
trajectories measured in the experiments. 
The fits are performed by minimizing the sum of the squared residuals, using 
the standard errors as weights in the objective function \cite{cordoba2022munch}.
At $70\%$ intensity of applied light, the value of the fitted parameter 
for the purely viscous model is $A=2.70 \times 10^{-12}~\text{N}$. 
The value of the parameter $A$ at $80\%$ intensity of the applied light 
is $A=4.01 \times 10^{-12}~\text{N}$. Finally, 
at $100\%$ intensity of the applied light, the value of the fitted parameter 
for the purely viscous model is $A=5.55 \times 10^{-12}~\text{N}$. Note that
the value of $A$ becomes about $1.5$ times larger when the intensity of the 
applied light is increased from $70\%$ to $80\%$. When the intensity 
of the applied light  is increased from $80\%$ to $100\%$, the value of $A$
increases by a factor of about $1.4$. \textcolor{black}{
These results are qualitatively consistent with the images in Figs \ref{active}F--H which also 
show an increasing size of the self-induced isotropic phase in 
the metal-coated side of the particles as the intensity of light is increased.}

\begin{figure}[t]
\center
\includegraphics[width=\linewidth]{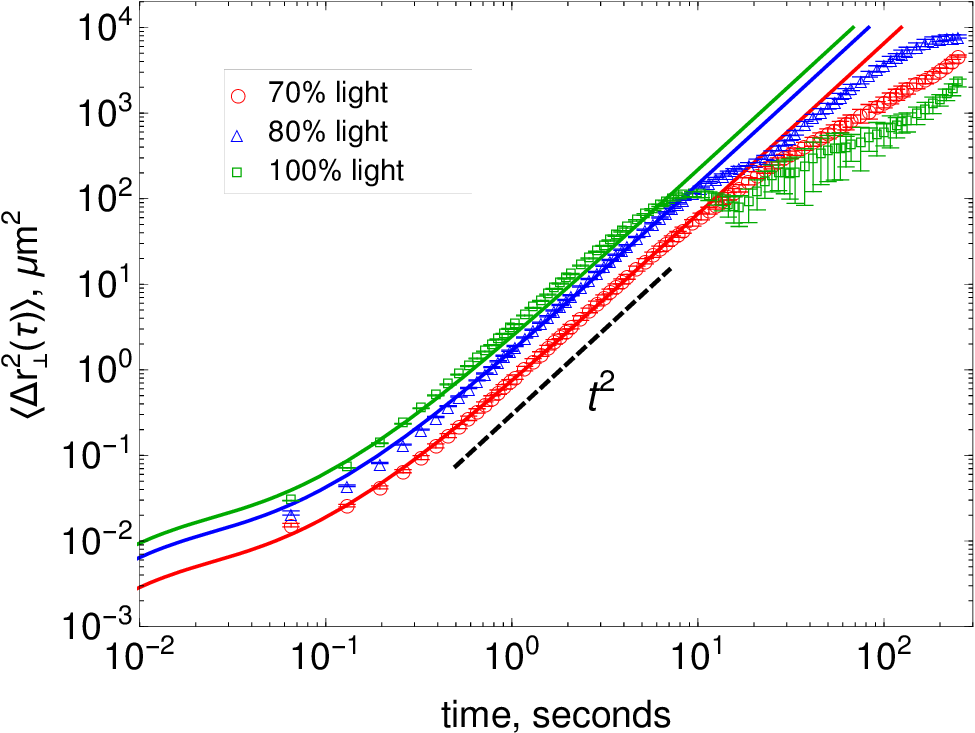}
\caption{Analysis of the mean-squared displacement of the Janus particle trajectories 
using the viscoelastic model.
The symbols are calculated using the particle tracking data and 
the continuous lines are fits of eq. (\ref{MSDfull}) to the data. 
The values of the fitted parameters for each intensity of the applied light are:
$A=30.73\times 10^{-12}~\text{N}$,
$K=458.43\times 10^{-6}$ N/m, ~~
$\lambda=0.076$ s (70\% intensity); 
$A=45.96\times 10^{-12}~\text{N}$, 
$K=458.43\times 10^{-6}$ N/m, ~~
$\lambda=0.076$ s (80\% intensity); and 
$A=55.72\times 10^{-12}~\text{N}$, 
$K=458.43\times 10^{-6}$ N/m, ~~
$\lambda=0.076$ s (100\% intensity).
Other parameters are set to $\ell_0=10~\mu\text{m}$, $\lambda_{\rm r}=1\mu\text{s}$,
$\zeta_{\rm r}=10 \zeta_{0,\perp}$.}
\label{fig2}
\end{figure}

Note also that the region of the $\langle \Delta r^2_{\perp} (\tau) \rangle$
that can be observed experimentally exhibits mostly ballistic behavior, 
\textit{i.e.} $\langle \Delta r^2_{\perp} (\tau) \rangle\sim \tau^2$.
However at short \update{lag-times} the region of the $\langle \Delta r^2_{\perp} (\tau) \rangle$ that 
can be observed in the experiments is limited by the tracking 
resolution, the size of the particles, and the viscosity of 
5CB at the temperature at which the experiments are performed. 
In the system considered here, times larger than $0.06$ s and 
mean-squared displacements larger than $0.01~\mu\text{m}^2$ can be observed.
Eq. (\ref{MSDvis}) indicates that for the system considered here
the diffusive part of $\langle \Delta r^2_{\perp} (\tau) \rangle$
dominates only at times shorter than about $10^{-3}$ s. Therefore, a sampling
frequency larger than 1000 Hz would be required in the particle tracking to observe 
that region. Moreover, given the very small displacements that occur at those 
time scales, a spatial resolution of about $10^{-6}~\mu$m would also be required.
These difficulties are typical of particle tracking in very viscous fluids such as 
nematic LCs. The advantage of active Janus particles like those 
employed here is that it is possible to observe motion without needing 
very high temporal and spatial resolution when tracking the particles.

Fig \ref{fig1} also shows that 
at the shorter times of the region where eq. (\ref{MSDvis}) 
is fitted to the experimental data, some significant deviations from the
ballistic behavior occur in the experimental $\langle \Delta r^2_{\perp} (\tau) \rangle$.
The motion of the Janus particle is clearly slower than ballistic at time 
scales ranging from $6\times10^{-2}$ s to $0.4$ s.
However eq. (\ref{MSDvis}) only predicts a transition towards motion 
slower than ballistic at times smaller than about $10^{-3}$ s.
This indicates that other physics not present in eq. (\ref{MSDvis}) may 
play a role in slowing the Janus particle motion at intermediate time scales. 
We hypothesize that the missing element in eq. (\ref{MSDvis}) is the 
viscoelasticity of the medium where the Janus particle is embedded. 
To account for that viscoelasticity, we have 
introduced a viscoelastic element in the friction tensor in eq. (\ref{zeta}).
The expression for the $\langle \Delta r^2_{\perp} (t) \rangle$ that results 
from this viscoelastic model is given in eq. (\ref{MSDfull}). When using this 
expression to analyze the experimental data, the parameters $A$, $K$, and $\lambda$
are used as fitting parameters. These parameters are related to the 
strength of the swimming force, the strength of the viscoelastic element, and 
the relaxation time of the viscoelastic element, respectively. 

Fig \ref{fig2} shows the fits of the viscoelastic model, eq. (\ref{MSDfull}), 
to the $\langle \Delta r^2_{\perp} (\tau) \rangle$ calculated from the particle
trajectories measured in the experiments. 
Again, the fits are performed by minimizing the sum of the squared residuals using 
the standard errors as weights in the objective function.
Note that these are the same 
$\langle \Delta r^2_{\perp} (\tau) \rangle$ data shown in Fig \ref{fig1} but we have 
shrunk the $x$ and $y-$axis scales to ease the evaluation of the quality of the 
fits to the experimental data. We find that a single set of 
values can describe the data at all 
intensities of the applied light for the parameters $K$ and $\lambda$. These are 
$K=458.43\times 10^{-6}~\text{N/m}$ and $\lambda=0.076~\text{s}$.
On the other hand, the fitted value of the parameter $A$ does change 
for different values of the intensity of applied light. 
At $70\%$ intensity of applied light, the value of the fitted parameter 
for the viscoelastic model is $A=30.73 \times 10^{-12}~\text{N}$. 
The value of the parameter $A$ at $80\%$ intensity of the applied light 
is $A=45.96 \times 10^{-12}~\text{N}$. And finally, 
at $100\%$ intensity of the applied light, the value of the fitted parameter 
for the viscoelastic model is $A=55.72 \times 10^{-12}~\text{N}$.
 Note that the quality of the 
fits using the viscoelastic model, eq. (\ref{MSDfull}), are better at the intermediate time 
scales, $4\times10^{-2}~\text{s} \lesssim t \lesssim 1~\text{s}$, at which the purely 
viscous model, eq. (\ref{MSDvis}) appeared to be failing. 
This indicates that the viscoelasticity of the medium
plays an important role in the dynamics of the Janus particles.
In Table \ref{tab:pars}, the values of the fitted parameters for 
both the purely viscous and the viscoelastic models are summarized. 

\begin{table}[h]
\centering
\renewcommand{\arraystretch}{0.8}
\addtolength{\tabcolsep}{3pt}    
  \caption{Summary of the values of the fitted parameters 
resulting from fitting eq. (\ref{MSDvis}) or eq. (\ref{MSDfull}) to the 
experimental data.}
  \label{tab:pars}
 \begin{tabular}{cccc}
\hline
Model & \multicolumn{3}{c}{Parameters} \\
& $A \times 10^{12}$ & $ K \times 10^{6}$ & 
$\lambda$\\ 
& N & N/m & s \\
\hline
\multicolumn{4}{c}{Light at 70\% intensity} \\
Purely viscous, eq. (\ref{MSDvis}) & $2.70$ & -- & -- \\
Viscoelastic, eq. (\ref{MSDfull}) & $30.73$ & $458.43$ & $0.076$ \\
\hline
\multicolumn{4}{c}{Light at 80\% intensity} \\
Purely viscous, eq. (\ref{MSDvis}) & $4.01$ & -- & -- \\
Viscoelastic, eq. (\ref{MSDfull}) & $45.96$ & $458.43$ & $0.076$ \\
\hline
\multicolumn{4}{c}{Light at 100\% intensity} \\
Purely viscous, eq. (\ref{MSDvis}) & $5.55$ & -- & -- \\
Viscoelastic, eq. (\ref{MSDfull}) & $55.72$ & $458.43$ & $0.076$ \\
\hline
\end{tabular}
\end{table}

It can be observed in Table \ref{tab:pars} 
that when the viscoelastic model is used in the analysis
the value of  $A$ becomes about 1.5 times larger when the intensity of the 
applied light is increased from $70\%$ to $80\%$.
When the intensity 
of the applied light  is increased from $80\%$ to $100\%$ the value of $A$
increases by a factor of about $1.2$.
This is similar to what was observed when the analysis was performed
with the purely viscous model, and again, 
\textcolor{black}{it indicates that increasing the intensity of 
the applied light produces a larger self-induced isotropic phase in 
the metal-coated side of the particles leading to higher swimming forces.}
Moreover, note that the viscoelastic model requires larger values of $A$ than the 
purely viscous model to fit the experimental data well.
This is not surprising since, in the viscoelastic model, the 
particle has to overcome the elasticity of the medium 
to achieve motion at a constant velocity (\textit{i.e.} ballistic motion).
\textcolor{black}{The results presented above
suggest that thermophoretic Janus particles may be used as 
micro-probes for measuring the rheological properties of a nematic LC. 
Such methodology would require first a 
systematic calibration of the swimming force as a function of 
the intensity of applied light. Once such calibration exists one can fix, 
for a given intensity of the applied light, the 
value of $A$ in the model equations and treat the 
friction parameter (\textit{e.g.}, $\zeta_{0,\perp}$) as adjustable when fitting 
the experimental  $\langle \Delta r^2_{\perp} (\tau) \rangle$. 
The Stokes relation for LCs, eq. (\ref{zeta0}),
may then be employed to extract rheological properties from the measured 
friction coefficients.}

\section{CONCLUSIONS} \label{conclusions}

We report the self-propulsion of light-activated Janus particles in a \update{thermotropic} nematic liquid crystal. 
The Janus particles used here are $10~\mu$m silica particles half-coated with titanium. The 
particles are embedded in nematic 4-Cyano-4'-pentylbiphenyl (5CB).
We observe that the Janus particles move toward their Ti-coated side and perpendicular to the nematic director, given the confinement of the colloidal-liquid crystals.
Based on this observation, we have proposed that the larger light absorption at the 
metal-coated side of the particle creates a local temperature gradient that \update{induces a nematic-isotropic phase transition that} drives the motion. The trajectories of the Janus particles in the nematic liquid crystal were tracked 
at different intensities of the applied light.
Previous reports exist of self-thermophoresis of Janus particles in isotropic, 
purely viscous fluids, but we believe this to be the first report of a
self-thermophoretic particle swimming in a nematic liquid crystal.

A model was proposed to describe the mean-squared displacement
of the Janus particles in the nematic liquid crystal. The model 
assumes that the Janus particle feels the liquid crystal as a continuum 
with the anisotropic viscosity of the bulk nematic phase. 
Since we propose that the active motion of the Janus particles is driven by self-thermophoresis, the swimming force
is assumed to be proportional to the local temperature gradient along the particle. The model
describes well the mean-squared displacement of the Janus particles in the ballistic region. 
The proposed model is fitted to the experimental data using the strength of the 
self-thermophoretic force as the fitting parameter. We find that 
the magnitude of the fitted parameter increases with the intensity 
of the applied light. This agrees with the hypothesis that 
a self-thermophoretic mechanism drives the swimming motion.
Moreover, we have also accounted for the 
viscoelasticity of the medium where the Janus particle is embedded. When this viscoelasticity is taken 
into account, the fits of the model to the experimental data improve, especially 
at shorter times.

The model proposed here does not describe the dynamics at long lag-times, the full particle trajectories during 5 min recordings reveal paths that include both straight and circular motion. We propose that these paths are the result of three main contributions: the non-uniform titanium coating that generates a non-homogeneous temperature increase on the particle's surface, the possible hydrodynamic interactions between particles, and we expect that the more important effect is the long-range inter-particle interactions driven by elastic distortions within the LC bulk. At long times the trajectory of the particles can be affected by the presence of other particles or impurities that create topological defects. The particles can feel the presence of other particles at distances of tens of the particle's diameter size. The study of the hydrodynamic and elastic inter-particle interactions lies out of the scope of this paper. Still, an investigation in this regard is in progress.

In nature, it is common for microswimmers, such as bacteria or eukaryotic cells, to swim 
in fluids where highly concentrated solutions of filamentous proteins such 
as actin or tubulin can form nematic phases. 
Hence, understanding the dynamics 
of microswimmers in nematic phases can have important applications  
in medicine, specifically in optimizing targeted delivery 
of pharmaceuticals. Here we have reported, we believe for the first time, 
a relatively simple synthetic system that can mimic some of the 
physics of those biological systems. Accordingly, the proposed system 
may serve as a tool for understanding some of the underlying physics of
the more complex biological systems.
Furthermore, the results presented here also suggest that thermophoretic Janus particles may be used as micro-probes for measuring the rheological properties of nematic liquid crystals. \textcolor{black}{Our investigation contributes to developing a platform for studying highly structured anisotropic active materials.}

\makeatletter
\def\@seccntformat#1{%
  \expandafter\ifx\csname c@#1\endcsname\c@section\else
  \csname the#1\endcsname\quad
  \fi}
\makeatother


\subsection*{AUTHOR DECLARATION}
The authors declare no competing interest.

\subsection*{ACKNOWLEDGEMENTS}
A. T.-V. thanks Prof. Erick Sarmiento-G\'{o}mez for the discussions and technical feedback on the experiments and data analysis, Prof. Teresa Lopez-Leon, Prof. Fransesc Sagu\'{e}s and Prof. Rui Zhang for the fruitful discussions and suggestions, and Dr. Zhengyang Liu and Lars K\"{u}rten for the suggestions on particle tracking. The authors gratefully acknowledge Prof. Stuart Rowan for sharing their polarized optical microscope that permitted us to perform the experiments. This work made use of the Searle Cleanroom at the University of Chicago, funded through Award Number C06RR028629 from the National Institutes of Health–National Center For Research Resources. Instrumentation was procured with funding generously provided by The Searle Funds at The Chicago Community Trust (Grant A2010-03222). The SEM technique was performed at the Pritzker Nanofabrication Facility of the Institute for Molecular Engineering at the University of Chicago, which receives support from Soft and Hybrid Nanotechnology Experimental (SHyNE) Resource (NSF ECCS-2025633), a node of the National Science Foundation’s National Nanotechnology Coordinated Infrastructure, RRID: SCR\_022955. \textcolor{black}{This work was also supported by the University of Chicago Materials Research Science and Engineering Center, which is funded by the National Science Foundation under award number DMR-2011854.}

\appendix

\section{APPENDIX A. DERIVATION OF THE MODEL EQUATIONS}\label{appA}

Here the mathematical details to obtain eq. (\ref{MSDfull}) are given.
To solve eq. (\ref {1_GLE}) for $\bm{r}_{\rm b}(t)$ we take the Laplace transform,
\begin{align}\label{1_GLEF}
-\bm{r}_{\rm b}(0)+m s^2 \bm {r}_{\rm b}[s]=
\frac{A}{\ell_0}\bm{u}_{\rm b}[s]-
s \bm{\zeta}[s] \cdot \bm {r}_{\rm b}[s]+
\bm{f}_{\rm B}[s].
\end{align}
Where $\bm{r}_{\rm b}(0)$ is the position of the Janus 
particle at the moment when the light source is turned on, 
$\bm{r}_{\rm b}[s]=\mathcal{L}\{\bm {r}_{\rm b}(t)\}$,
$\bm{u}_{\rm b}[s]=\mathcal{L}\{\bm{u}_{\rm b}(t)\}$ and 
$\bm{f}_{\rm B}[s]=\mathcal{L}\{\bm {f}_{\rm B}(t)\}$ are the Laplace transforms 
of $\bm{r}_{\rm b}(t)$, $\bm{u}_{\rm b}(t)$ and $\bm{f}_{\rm B}(t)$, respectively.
The Laplace transform of $\bm{\zeta}(t)$ is given by,
\begin{align}
\bm{\zeta}[s]=\bm{\zeta}_0+\bm{\delta} \frac{K \lambda}{1+s\lambda}.
\end{align}
Neglecting particle inertia, \textit{i.e.} $m\approx0$, 
setting the origin of the coordinate system 
at the initial position of the particle, $\bm{r}_{\rm b}(0)=\bm{0}$, 
and solving eq. (\ref{1_GLEF}) for $\bm {r}_{\rm b}[s]$ gives,
\begin{align}
\bm {r}_{\rm b}[s]=\frac{\bm{\zeta}[s]^{-1}}{s}\cdot\left(\frac{A}{\ell_0}\bm{u}_{\rm b}[s]+
\bm{f}_{\rm B}[s]\right).
\end{align}

To obtain $\bm{u}_{\rm b}[s]$, we solve eq. (\ref{ecQ1}) by again taking the Laplace transform,
\begin{eqnarray}\label{ecQ2}
s \bm{u}_{\rm b}[s]=-\frac{1}{\lambda_{\rm r}} \left( 1-\frac{\ell_0}{u_{\rm b}}\right) \bm{u}_{\rm b}[s]+\bm{g}_{\rm B}[s].
\end{eqnarray}
The spring term is nonlinear in $\bm{u}_{\rm b}$; therefore to be able to 
solve for $\bm{u}_{\rm b}[s]$ analytically we take 
only the linear term of a Taylor series expansion around $\ell_0$,
\begin{align}\label{ecQ3}
-\ell_0+s u_{{\rm b},\perp}[s]=& \frac{1}{\lambda_{\rm r}} \left( \frac{\ell_0}{s} - 
u_{{\rm b},\perp}[s] \right)+g_{{\rm B},\perp}.
\end{align}
Where $u_{{\rm b},\perp}$ is the component of $\bm{u}_{\rm b}$ perpendicular to the nematic director 
$\hat{\bm{n}}$. And we have assumed that $\bm{u}_{\rm b}$ is initially perpendicular to $\hat{\bm{n}}$
and does not deviate too much from this direction during the course of the experiment.
We also assume that $u_{\rm b}$ does not significantly deviate from $\ell_0$.
This means that the analytic model derived here is valid for $\zeta_{\rm r}\gg\zeta_\perp$ and
$t\gg\lambda_{\rm r}$ only.
Solving for $u_{{\rm b},\perp}$ in eq. (\ref{ecQ3}) one gets,
\begin{align}\label{eqQsol}
u_{{\rm b},\perp}=\frac{\ell_0}{s}+\left(\frac{\lambda_{\rm r}}{1+\lambda_{\rm r} s}\right) g_{{\rm B},\perp}[s].
\end{align}

In experiments  the component of $\bm {r}_{\rm b}[s]$ perpendicular to $\hat{\bm{n}}$  is
tracked. This component is given by,
\begin{align}
r_{{\rm b},\perp}[s]=\frac{1}{s \zeta_{0,\perp}[s]}
\left(\frac{A}{\ell_0}u_{{\rm b},\perp}[s]+
f_{{\rm B},\perp}[s]\right).
\end{align}
We now calculate the mean-squared displacement (MSD) from the particle trajectories,
$\langle \Delta r_{{\rm b},\perp}^2 (t)\rangle=
\langle [r_{{\rm b},\perp}(t)-r_{{\rm b},\perp}(0)]^2\rangle$ and
since we have assumed $r_{{\rm b},\perp}(0)=0$ then 
$\langle \Delta r_{{\rm b},\perp}^2 (t)\rangle=
\langle r_{{\rm b},\perp}(t)^2\rangle$.
To obtain $\langle r_{{\rm b},\perp}(t)^2\rangle$ we start by calculating
$\left\langle r_{{\rm b},\perp}[s] r_{{\rm b},\perp}[s']\right\rangle$,
\begin{align}\label{rcorr}
\nonumber
\left\langle r_{{\rm b},\perp}[s] r_{{\rm b},\perp}[s']\right\rangle&=
\frac{1}{s s' \zeta_\perp[s]\zeta_\perp[s']}
\bigg(\frac{A^2}{\ell_0^2} \left \langle u_{{\rm b},\perp}[s]u_{{\rm b},\perp}[s'] \right\rangle \\ &+
\left\langle f_{{\rm B},\perp}[s] f_{{\rm B},\perp}[s']\right\rangle_{\rm eq}\bigg),
\end{align}
where we have used $\langle f_{{\rm B},\perp}[s]\rangle_{\rm eq}=0$
and we will use the FDT to find $\left\langle f_{{\rm B},\perp}[s]
f_{{\rm B},\perp}[s']\right\rangle_{\rm eq}$,
\begin{align} 
\nonumber
\left\langle f_{{\rm B},\perp}[s]
f_{{\rm B},\perp}[s']\right\rangle_{\rm eq}=&k_{\rm B}T
\int_{0}^{\infty}\int_{0}^{\infty}\zeta_{\perp}(t-t')e^{-s t-s' t'}dt dt' \\ \nonumber
=&k_{\rm B}T
\int_{0}^{\infty}\int_{0}^{\infty}\bigg[\delta(t-t')  \zeta_{0,\perp}+ \\ \nonumber
&H(t-t')K e^{-\lambda/t}\bigg]e^{-s t-s' t'}dt dt' \\
=&k_{\rm B}T  \left[\frac{\zeta_{0,\perp}}{s+s'´} +
\frac{K}{(s+s')(s+1/\lambda)}\right].
\end{align}

The term involving $u_{{\rm b},\perp}$ in eq. (\ref{rcorr}) can 
be obtained from eq. (\ref{eqQsol}),
\begin{align}\label{qcorr}
\left \langle u_{{\rm b},\perp}[s]u_{{\rm b},\perp}[s'] \right\rangle=
\frac{\ell_0^2}{s s'}+
\frac{k_{\rm B} T \lambda_{\rm r}^2}{(s+s')\left(1+s\lambda_{\rm r} \right)
\left(1+s'\lambda_{\rm r} \right) \zeta_{\rm r}}.
\end{align}
The FDT for $\bm{g}_{\rm B}(t)$, eq. (\ref{fdtg}), was used when 
writing eq. (\ref{qcorr}),
\begin{align}
\nonumber
\left\langle g_{{\rm B},\perp}[s]
g_{{\rm B},\perp}[s']\right\rangle_{\rm eq}=&k_{\rm B}T
\int_{0}^{\infty}\int_{0}^{\infty}\frac{\delta(t-t')}{\zeta_{\rm r}} dtdt' \\
=&\frac{k_{\rm B}T}{(s+s')\zeta_{\rm r}}.
\end{align}

In what follows we will assume $s\lambda_{\rm r}\ll1$ which is equivalent to
assuming $t/\lambda_{\rm r}\gg1$. This means the model is appropriate only when 
the dumbbell behaves as a rigid object and therefore $\lambda_{\rm r}$
should always be chosen to be smaller than all the other time scales in the model.
This is consistent with the assumptions that were used to derive the expression for 
$u_{{\rm b},\perp}[s]$ in eq. (\ref{ecQ2}). 
With those assumptions, one can obtain,
\begin{align}
\nonumber
&\left\langle r_{{\rm b},\perp}[s] r_{{\rm b},\perp}[s']\right\rangle \\
& =
\frac{k_{\rm B}T  \left[\frac{\zeta_{0,\perp}}{s+s'´} +
\frac{K}{(s+s')(s+1/\lambda)}\right]+A^2\left[ \frac{\ell_0^2}{s s'}+
\frac{k_{\rm B} T \lambda_{\rm r}^2}{(s+s') \zeta_{\rm r}}\right]}
{s s'\left(\zeta_{0,\perp}+\frac{K\lambda}{1+s\lambda}\right)
\left(\zeta_{0,\perp}+\frac{K\lambda}{1+s'\lambda}\right)}.
\end{align}
The next step to obtain $\langle r_{{\rm b},\perp}(t)^2\rangle$ is
to take the inverse Laplace transform twice,
\begin{align}
\left\langle r_{{\rm b},\perp}[t] r_{{\rm b},\perp}(t')\right\rangle=
\mathcal{L}^{-1} \left\{ \mathcal{L}^{-1} \left\{ \left\langle r_{{\rm b},\perp}[s] r_{{\rm b},\perp}[s']
\right\rangle\right\}\right\},
\end{align}
and finally, by making $t'=t$, one obtains eq. (\ref{MSDfull}).
\textcolor{black}{By taking the limit $K\longrightarrow0$ in eq. (\ref{MSDfull})
a purely viscous model can be recovered. The expression for the 
$\langle \Delta r_{{\rm b},\perp}^2(t)\rangle$ in that limit is given in 
eq. (\ref{MSDvis}). Another limiting case that may be illustrative occurs 
when $t\gg\lambda$,
\begin{align}\label{MSDlt}
\nonumber
\langle \Delta r_{{\rm b},\perp}^2(t)\rangle&=
k_{\rm B} T \left( \frac{1}{\zeta_{0,\perp} + K \lambda} + 
\frac{A^2 \lambda_{\rm r}^2}{\ell_0^2\zeta_{\rm r} (\zeta_{0,\perp}+K\lambda)^2}\right)t\\
&+ \left(\frac{A}{\zeta_{0,\perp}+K \lambda} \right)^2 t^2.
\end{align}
Note that eq. (\ref{MSDlt}) is equal to eq. (\ref{MSDvis}) in the limit $K\longrightarrow0$ 
as it should be.}

\bibliography{referencesLC,referencesswimmer,referencesRev} 
\section*{SUPPORTING INFORMATION DESCRIPTION}

{\parindent0pt
Request supporting files directly to\\
\textcolor{blue}{tavera@uchicago.edu} and \textcolor{blue}{atv.tavera@gmail.com}
\\
Here is a description of Movies S1 to S6.
\\
\\
\textbf{MovieS1 (separate file)} \\
Cross-polarized movie of activated Janus particles with $70\%$ light intensity. The birefringent response is noticeable due to the liquid crystal anchoring on the particles’ surface. At $t=0$, the light is off and turned on within $t=3$ s. The total duration of the experiment was about $17$ min. The movie was sped up.
\\
\\
\textbf{MovieS2 (separate file)}\\
Bright-field movie of an activated Janus particle with $70\%$ light intensity. The particle undergoes a random trajectory shown as a colored path. The trajectory is initially straight, and eventually deflects to become spiral-like. The total duration of the experiment was 5 min. The movie was sped up.
\\
\\
\textbf{MovieS3 (separate file)}\\ 
Bright-field movie of an activated Janus particle with $80\%$ light intensity. The particle undergoes a random spiral-like trajectory shown as a colored path. The total duration of the experiment was $5$ min. The movie was sped up.
\\
\\
\textbf{MovieS4 (separate file)}\\ 
Bright-field movie of an activated Janus particle with $100\%$ light intensity. The particle undergoes a random trajectory shown as a colored path. The trajectory is primarily straight, with a pronounced deflection at the end. The total duration of the experiment was $5$ min. The movie was sped up.
\\
\\
\\
\\
\\
\\
\\
\\
\\
\\
\\
\\
\\
\\
\\
\\
\\
\\
\\
\\
\\
\\
\\
\\
\\
\\
\\
\\
\\
\\
\\
\textbf{MovieS5 (separate file)}\\
Bright-field movie of activated Janus particles. The movie begins with the light off, then turns on at $70\%$, and finishes with the light turned off. Two particles undergo different random trajectories. Two other Janus particles stay attached to the substrate but also self-induce a nematic-isotropic transition. The paths taken by the motile particles are influenced by the presence of the immotile particles. One of the particles ended up colliding with one of the static particles. The other motile particle goes out of the field of view and returns to keep a random trajectory. The total duration of the experiment was about $17$ min. The movie was sped up.
\\
\\
\textbf{MovieS6 (separate file)}\\ 
Bright-field movie of activated Janus particles. The movie begins with the light off, then turns on at $80\%$. Two particles undergo different random trajectories. Three other Janus particles stay attached to the substrate but also self-induce a nematic-isotropic transition. The paths taken by the motile particles are influenced by the presence of the immotile particles and impurities in the sample. One of the motile particles is trapped by one of the impurities. The other motile particle keeps a random trajectory, avoiding the contact with the other particles and impurities. The total duration of the experiment was about $8$ min. The movie was sped up.

}

\end{document}